\documentclass[preprint,amsmath,amssymb,aps,prb,superscriptaddress]{revtex4-1}

\usepackage{graphicx}
\usepackage{dcolumn}
\usepackage{bm}

\begin{document}

\title{Formation of the $n=0$ Landau level in hybrid graphene}

\author{P. Cadden-Zimansky}
\email{paulcz@bard.edu}
\affiliation{Physics Program, Bard College, Annandale-on-Hudson, New York 12504, USA}

\author{M. Shinn}
\affiliation{Physics Program, Bard College, Annandale-on-Hudson, New York 12504, USA}
\affiliation{Department of Physics, Washington University, St. Louis, Missouri 63130, USA}

\author{G. T. Myers}
\affiliation{Physics Program, Bard College, Annandale-on-Hudson, New York 12504, USA}
\affiliation{Department of Physics, Boston College, Chestnut Hill, Massachusetts 02467, USA}

\author{Y. Chu}
\affiliation{Physics Program, Bard College, Annandale-on-Hudson, New York 12504, USA}
\affiliation{Department of Mathematics, Duke University, Durham, North Carolina 27708, USA}

\author{M. J. Dalrymple}
\affiliation{Physics Program, Bard College, Annandale-on-Hudson, New York 12504, USA}

\author{H. C. Travaglini}
\affiliation{Physics Program, Bard College, Annandale-on-Hudson, New York 12504, USA}
\affiliation{Department of Physics, University of California, Davis, California 95616, USA}

\date{\today}

\begin{abstract}
The minimum of 4-terminal conductance occurring at its charge neutral point has proven to be a robust empirical feature of graphene, persisting with changes to temperature, applied magnetic field, substrate, and layer thickness, though the theoretical mechanisms involved in transport about this point -- vanishing density of states, conventional band gap opening, and broken symmetry quantum Hall mobility gaps -- vary widely depending on the regime.  In this paper, we report on observations of a regime where the 4-terminal conductance minimum ceases to exist:  transport in monolayer graphene connected to bilayer graphene during the onset of the quantum Hall effect.  As monolayer and bilayer graphene have distinct zero-energy Landau levels that form about the charge neutral point, our observations suggest that competitions between the differing many-body orderings of these states as they emerge may underlie this anomalous conductance. 
\end{abstract}

\pacs{72.80.Vp, 73.22.Pr, 73.43.-f}
\maketitle

With an improving understanding of topological phases -- electronic states of matter characterized by their topological order rather than their symmetries -- attention is turning to how these phases interact with other electronic orders and how two distinct topological phases interact with each other.    Such topological heterostructures hold the promise of reliably realizing diverse new phenomena, such as magnetic monopoles,\cite{qi} Majorana fermions,\cite{fu} and topological Hall effects.\cite{yasuda}  However, theoretical modeling of many heterostructure devices posits a clean interface between the constituent parts across which phase coherence is preserved,\cite{lutchyn} a requirement that can often be challenging to realize experimentally in fabricating heterostructures.

Hybrid graphene, graphene heterostructures consisting of regions of differing atomic thicknesses, offers perhaps the cleanest possible system to study competition between differing topological states.  The pristine interface of these structures is achieved by extracting them from a single graphite crystal, while the differing topological states are a result of graphene layers of differing thicknesses supporting distinct quantum Hall states under the application of a magnetic field.  More specifically, the charge carriers in monolayer graphene have a $\pi$ Berry phase and form a four-fold degenerate $n=0$ Landau level that straddles the electron-hole, charge neutral (``Dirac'') point,\cite{novo2, zhang1} while the $2\pi$ Berry phase carriers in bilayer graphene form a corresponding eight-fold degenerate Landau level about the same point.\cite{novo3}

Thus far, relatively few studies have examined the competing quantum Hall states in hybrid monolayer-bilayer graphene,\cite{puls, tsukuda, tian, Iqbal, yan, hu, zhao} with those experimental studies that have been done mainly focusing on effects due to transport across the hybrid interface.\cite{puls, tsukuda, tian, Iqbal}  In contrast, we here report on conventional longitudinal and transverse resistance measurements using electrical contacts only on the bulk graphene monolayer part of hybrid structures.  While many of the measurements display quantum Hall formation and states identical to homogenous monolayer graphene, our results show that the presence of a bilayer perturbation gives rise to qualitatively distinct transport features in the monolayer during the formation of the $n=0$ Landau level.  These features are potentially indicative of a new type of electronic ordering in hybrid graphene that is not found in homogenous graphene.

\begin{figure}
\includegraphics[width=8cm]{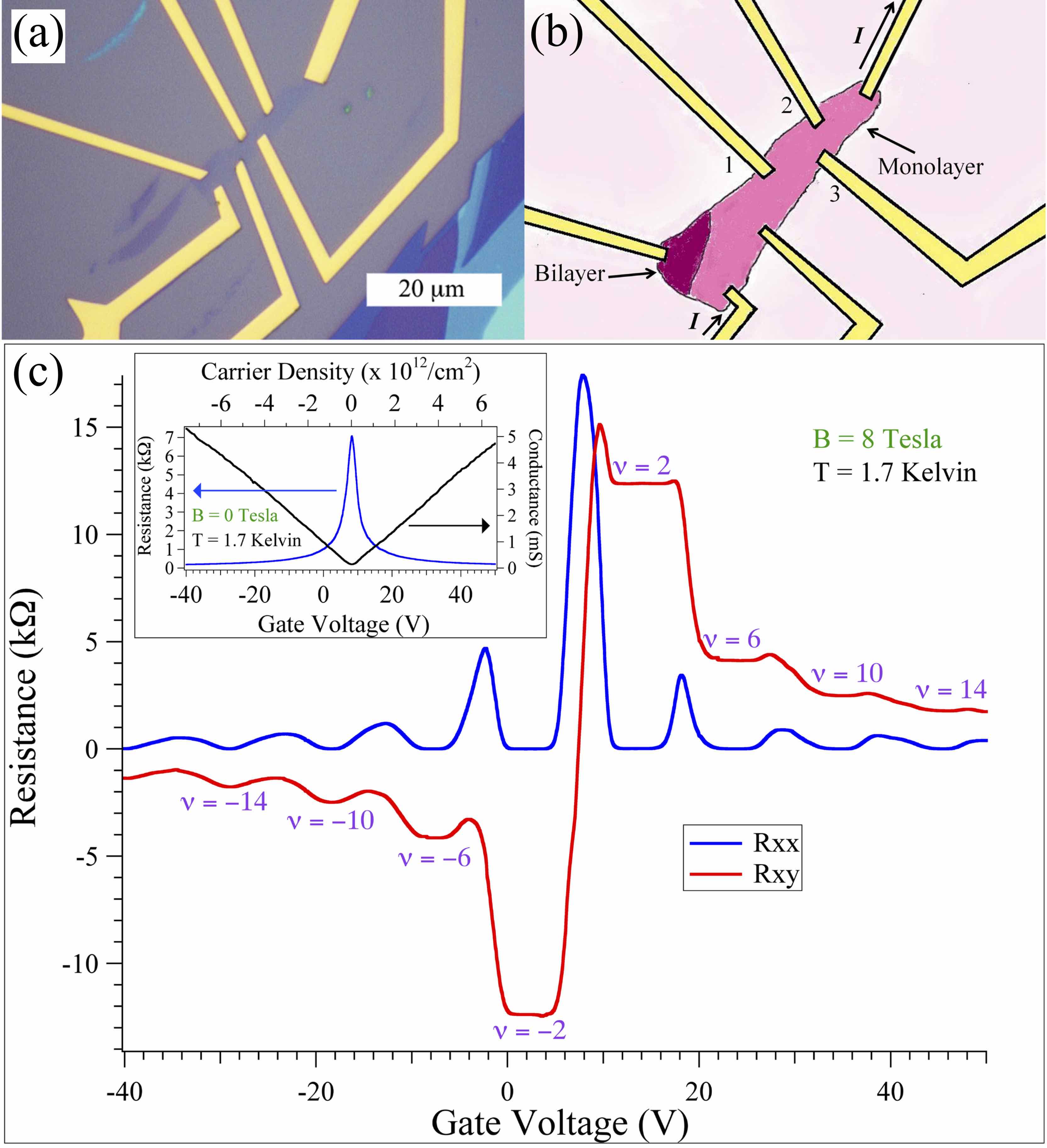}
\caption{\label{Fig1} (a) Optical image of hybrid graphene Sample A.  (b) Schematic showing the measurement configuration of the sample.  Longitudinal resistance R$_{xx}$ is measured from leads 1 \&  2, transverse R$_{xy}$ is measured between leads 3 \& 2.  The lead on the bilayer section of the graphene is not used for these measurements.  (c) R$_{xx}$ and R$_{xy}$ measurements at 8 T and 1.7 K.  The transverse R$_{xy}$ plateaus correspond to labeled filling factors $\nu$ and are identical to the quantum Hall filling sequence found in homogenous monolayer graphene.  The inset shows the 0-field resistance and conductance.}
\end{figure}

Measurements were conducted on two hybrid graphene flakes, Sample A and Sample B, shown in the first panels of Figs.~1 and 2 respectively.   Each flake is fabricated by mechanical exfoliation of graphite on a SiO$_2$/doped Si substrate where regions of varying thickness are identified by optical contrast.  Larger regions of bilayer graphene for each sample were etched away using an electron-beam patterned PMMA mask and O$_2^+$ plasma to leave a predominantly monolayer flake (60 $\mu$m$^2$ for sample A, 100 $\mu$m$^2$ for sample B) with a small bilayer region at one corner (5 $\mu$m$^2$ for sample A, 10 $\mu$m$^2$ for sample B).  1 nm/80 nm Cr/Au leads were patterned and deposited using electron-beam lithography and thermal evaporation.  While sample A has one lead on its bilayer section, all measurements presented here used only the leads attached to the monolayer portions of the flakes, as shown in the measurement configurations of the upper right panels of Figs.~1 and 2.

Transport data was collected using conventional low-frequency (9-17 Hz), lock-in techniques using currents ranging from 10 nA to 100 nA.  4-terminal longitudinal (R$_{xx}$) and transverse (R$_{xy}$) resistances were measured at the base temperatures of a $^4$He cryostat (1.7 K, sample A) and $^3$He cryostat (0.35 K, sample B), while the carrier density of the graphene was adjusted using a gate voltage (V$_g$) applied to the conducting Si substrate.  We correlate the carrier density to the gate voltage using low-field Hall measurements, and from measurements of the longitudinal sheet resistance $R_{s}$ and $R_{xy}$ find the mobility $\mu=\frac{1}{R_s}\frac{\partial{R_{xy}}}{\partial{B}}$ of the monolayer sections at a carrier sheet density $n_s=2 \times 10^{12}$ cm$^{-2}$ to be 6,500 [6,400] cm$^2$/V$\cdot$s for sample A and 6,900 [5,600] cm$^2$/V$\cdot$s for sample B for the electron [hole] bands.

Fig.~1c shows R$_{xx}$ and R$_{xy}$ data as a function of gate voltage for sample A subjected to a perpendicular magnetic field of 8T, which are entirely consistent with the integer quantum Hall effect for monolayer graphene.\cite{novo2, zhang1}  The transverse resistance plateaus R$_{xy}\,$=$\,h/\nu{e^2}$ occur at filling factors $\nu\,$=$\,\ldots-10, -6, -2, 2, 6, 10 \ldots$ reflecting the 4-fold spin-valley degeneracy of each Landau level (indexed by $n\,$=$\,\ldots -2, -1, 0, 1, 2 \ldots$) and in contrast to the homogenous bilayer filling sequence $\nu\,$=$\,\ldots-12, -8, -4, 4, 8, 12\ldots$ that demarcate an 8-fold degenerate Landau level about its charge neutrality point.\cite{novo3} We note, what often goes unremarked on in graphene quantum Hall data, that the resistance maxima at the half filling of each Landau level, between the R$_{xx}$ zeros, outline a peak about the charge neutrality point that mirrors the 0-field resistance peak shown in the inset.  

Focusing on the maximum at the charge neutral point, this ``Dirac peak'' has proven to be a robust empirical feature at all fields in both homogenous monolayer graphene and homogenous bilayer graphene.\cite{checkelsky1, feldman}  The physical mechanisms underlying this peak, however, can depend on the sample and field range.  For monolayer graphene at 0 field, the linear dispersion relation leading to a vanishing density of states at the charge neutral point predicts a resistance peak, though the value of the conductivity at this point is complicated by considerations of disorder, substrate interactions, and inhomogeneity in samples.\cite{martin, mayorov, fuhrer}  In bilayer graphene asymmetries can cause a small band gap to open at this point.\cite{zhang3}  As the field is increased, localization effects can increase or decrease the peak resistance.\cite{morozov}  In the quantum Hall regime, the point corresponds to the half-filled $n=0$ state, where single-particle interactions with Landau levels affect the transport properties.\cite{abanin}  At high fields, typically $\geq$15 T for samples with mobilities comparable to ours, degeneracy breaking of the Landau level leads to an unusual $\nu=0$ filling factor at this point and a corresponding increase in the measured resistance.\cite{zhang2} To the authors' knowledge, across all these field regimes in homogenous graphene, the 4-terminal conductance at this point remains a minimum with respect to filling and gate voltage.

\begin{figure}
\includegraphics[width=8cm]{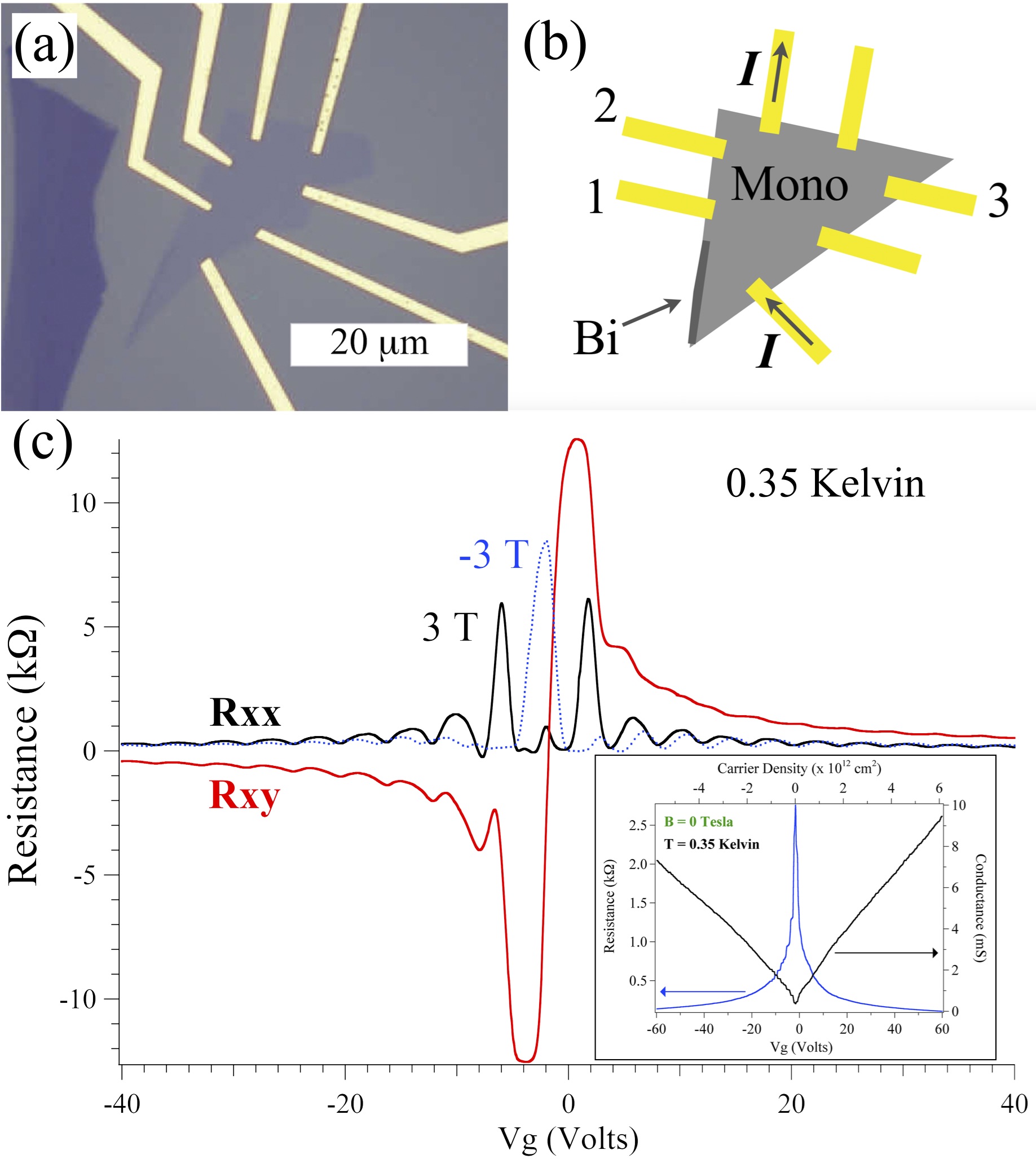}
\caption{\label{Fig2}(a) Optical image of hybrid graphene Sample B.  (b) Schematic showing the measurement configuration of the sample.  R$_{xx}$ is measured from leads 1 \&  2, R$_{xy}$ is measured between leads 3 \& 2.   (c) R$_{xx}$ and R$_{xy}$ measurements at $\pm$3 T and 0.35 K.  The R$_{xy}$ data shows plateaus forming at filling factors consistent with homogenous monolayer quantum Hall states while corresponding zeros in the R$_{xx}$ data form.  In contrast to homogenous monolayer, the Dirac peak is suppressed for one field polarity while the conductance of other half-filled Landau levels increases in comparison to the opposite polarity.  The inset shows the 0-field resistance and conductance of sample B.}
\end{figure}

\begin{figure}
\includegraphics[width=14cm]{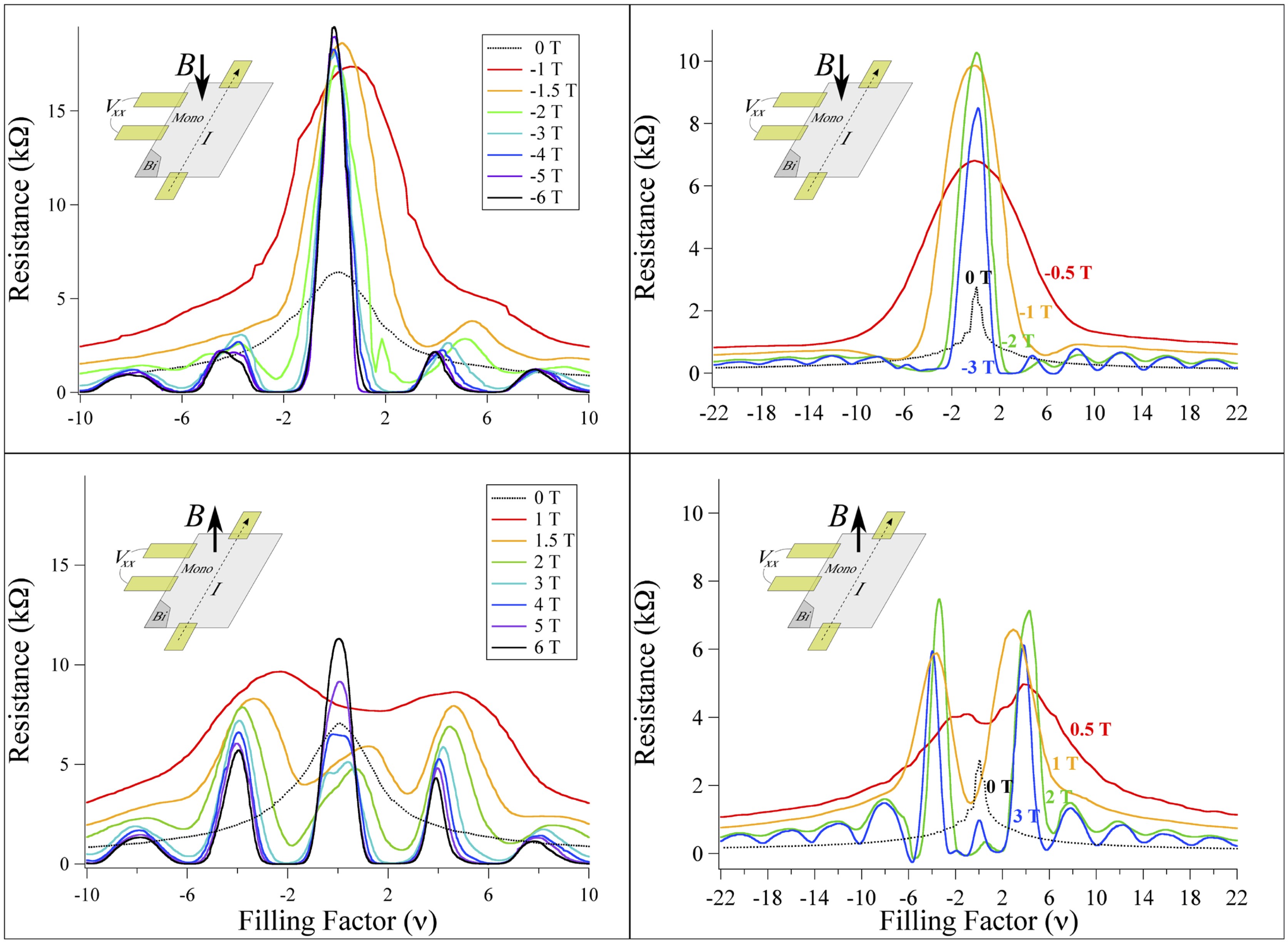}
\caption{\label{fig:wide}R$_{xx}=$V$_{xx}/$I of Sample A (left panels) and Sample B (right panels) for negative magnetic field (top panels) and positive magnetic field (bottom panels) as a function of Landau level filling.  The measurement configuration is shown schematically in the insets.  All panels show the onset of the resistance minima indicative of filled Landau levels and ballistic edge state conduction as the field is increased.  The top panels show an increase of the resistance at the charge neutral point up to a roughly constant value, corresponding to the 4-fold degenerate $n=0$ Landau level at  half-filling and consistent with previously reported field evolution of this point in homogenous monolayer graphene.  Conversely, resistance at the charge neutral point for the positive field does not monotonically increase, but rather displays an increase in conduction at the onset of the $n=0$ Landau level formation.  Note that since the zero-field data has no defined filling factors, the $x$-axis scaling for these traces is arbitrary.}
\end{figure}

In marked contrast, the monolayer sections of both hybrid graphene samples we measure show the charge neutral point can cease to be a conductance minimum.  Fig.~2c shows the longitudinal and transverse resistance at the onset of the quantum Hall effect for sample B.  While the R$_{xy}$ plateaus and R$_{xx}$ zeros of this regime are developing, the resistance at the charge neutral point is markedly different depending on the direction of the applied magnetic field.  A -3 T field reveals a Dirac peak similar to the zero-field and high-field regimes, while a field of opposite polarity suppresses the peak almost entirely.

The evolution of the nascent quantum Hall states with field for each sample is shown in Fig.~3.  The negative field evolution is qualitatively similar to the evolution observed for homogenous monolayer graphene; in particular, the Dirac peak resistance increases with field and settles at a roughly constant value as the quantum Hall resistance minima develop.  In stark contrast, a low, positive field causes the peak to ``break,'' becoming a local minimum rather than maximum, forming what might be termed a ``Dirac caldera.''  As the field is increased the peak re-emerges, though initially at a lower value than both the adjacent peaks corresponding to the half-filled $n=\pm1$ Landau levels and the positive field peak.  The suppression of the Dirac peak is more pronounced in the sample measured at lower temperature, where it almost disappears entirely at 2 T.

We emphasize that while this newly developed resistance minimum at the charge neutral point coincides with a field range where the resistance minima of filled Landau levels are emerging, there is evidence against there being topologically-protected, ballistic edge-state conduction at this point.  When samples are biased at a field and filling corresponding to an emerging quantum Hall state, the incompressible nature of the bulk states results in the R$_{xx}$ voltage probes becoming capacitively coupled across the insulating bulk.  As the state forms this capacitive coupling drives the measured AC voltage out of phase with the applied current, an effect that is observed for all the conventional degenerate monolayer fillings during Landau level formation.  Though there is a region in each sample where the resistance at the charge neutrality point decreases with increasing field, the voltage remains in phase with the current, implying it is conduction paths through the bulk that are being altered.

In addition to the conductance asymmetry with field at half-filling of the $n=0$ level, a conductance asymmetry at half-filling of the $n=\pm1$ levels can also be seen in Fig.~3.  For these levels it is the negative field resistance that is markedly lower than the positive field resistance.  While this asymmetry persists at higher Landau levels, as seen in Fig.~2c, it is less pronounced the farther away the sample is biased from the $n=0$ level.  Note also from Fig.~2c that the R$_{xx}$ peaks and minima are slightly more compressed as a function of gate voltage at positive field, an effect that also attenuates farther away from the charge neutral point.

We turn to the question of potential causes for the field asymmetry and Dirac peak suppression in these samples.  For a fixed current, one factor that could precipitate a measured voltage drop on one side of a sample is certain non-uniform current distributions.  As the counter-propagating hole and electrons currents are, by definition, equal at the charge neutral point, there exists a purely classical mechanism that can drive nonuniformity.  In the classical Hall effect the transverse forces on charge carriers are presumed to disappear as the transverse electric field $\vec{E}$ due to accumulated edge charge balances the $\vec{v}\times\vec{B}$ effect from the applied magnetic field.  At the charge neutral point the transverse $\vec{E}$ is measurably zero, so that the force from the magnetic field is unbalanced and can drive current away from our measurement contacts in the setups where the Dirac caldera is observed.  While this imbalance may play a role the observed effects, the facts that this imbalance would also be present in homogenous samples and that the anomalies are observed well outside the region of zero transverse electric field argue against it being their sole cause.

A quantum Hall consideration that may play a role is the posited Landau level levitation that occurs around the onset of the quantum Hall effect.  Hypothesized by Khmelnitskii and Laughlin shortly after the discovery of the integer quantum Hall effect to address the issue of what happens to the extended states responsible for quantized conduction as $B\to0$,\cite{khmelnitskii, laughlin} the theory states that the energies of the extended states levitate upwards in energy as the field is reduced according to $E_n(B)=\epsilon(n)(1+(\omega_c\tau)^{-2})$, where $\omega_c=eB/m$ is the cyclotron frequency, $\tau$ is the mean free scattering time, and $\epsilon(n)=(n+1/2)\hbar\omega_c$ are the high-field Landau level energies for conventional 2DEGs.  This theory yields a cross-over into the quantum Hall regime at $\omega_c\tau=1$, corresponding semi-classically to the region where the scattering length becomes comparable to the magnetic length and phase-coherent Landau level orbits can be established.  The $\omega_c\tau=1$ condition is equivalent to a $B=1/\mu$ condition, which for our samples is just under 2 T, the point at which the half-filling conductance asymmetries are at a maximum (Fig.~4a and b).

\begin{figure}
\includegraphics[width=8cm]{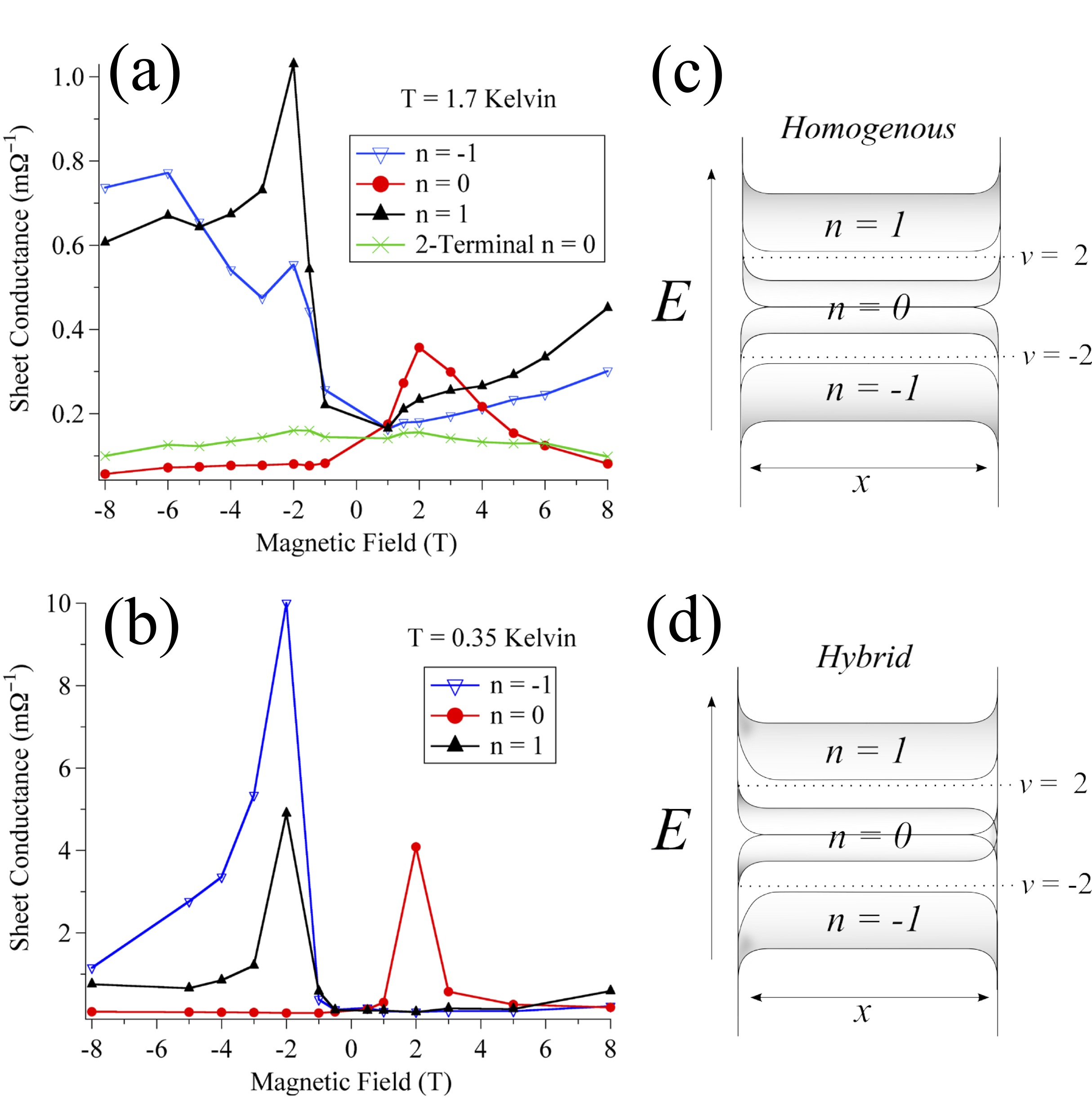}
\caption{\label{fig:wide}Sheet conductance of (a) sample A and (b) sample B at half-filling of the $n=0$ and $n=\pm1$ Landau levels.  The symmetric two-terminal conductance of sample A is also shown in (a).  All asymmetric anomalies show peaks around 2 T $\approx1/\mu$. (c) Schematic of lateral bending of levels at the sample edges causing quantized edge-state conductance; the particle-hole symmetric $n=0$ level has its 4-fold-degeneracy split at the edges.  The shading of each level reflects the single-particle, 0-field density of states. (d) Possible lateral bending with bilayer perturbation.  The alternate bending of the $n=0$ state at each end provides a mechanism for the observed asymmetry of the 4-terminal conductance at each edge when this level is half filled.  Coupled with level repulsion, this bending also provides an explanation for the asymmetric conductance of the half-filled $n=\pm1$ levels.}
\end{figure}

Since its introduction, the levitation idea even in conventional 2DEGs has continued to be subject to theoretical revision and experimental test.\cite{haldane, pan}  Potential effects due to levitation for the anomalous quantum Hall effect in graphene, where the high-field energies are given by $\epsilon(n)=sgn(n)\sqrt{2\hbar{c_g}^2eBn}$ ($c_g$ is the carrier speed), have just begun to be studied.\cite{huang} The zero-energy Landau level manifestly presents a paradoxical challenge to the original theory since it floats neither up nor down as the field is decreased.  Further complicating the accounting for how this level manifests itself at the cross-over field, the extended states of this level also bend in an unusual fashion spatially due to the confining edge potentials in order to produce the $\nu=\pm2$ filling factors (Fig.~4c).

While the bilayer perturbation adds an additional factor, the fact that both bilayer and monolayer have zero-energy Landau levels with different degeneracies that break differently at high fields may provide a mechanism for the observed conductance asymmetries.\cite{young, feldman}  If the competition between orderings led to the Fig.~4d alternate lateral bending within the $n=0$ level on each side of the sample, where the side is determined through the combination of the field direction and bilayer region placement, several empirical observations would be explained.  First, the crossing of states within this level on one side would lead to a conductance increase on this side at the charge neutral point.  Second, coupled with level repulsion, the alternate bending could lead to an asymmetric compacting of extended states in the adjacent levels, accounting for the conductance increase within the $n=\pm1$ states on the opposite side.  This compacting also predicts that these states would be pushed away from the $n=0$ level energetically on this side, consistent with the higher gate voltages required to start filling them in the transport data.

The theoretical details of how such a crossing could be precipitated by the bilayer region are in need of development.  On the empirical side, further characterization of the charge neutral anomaly with temperature, applied field direction, and higher mobility samples on boron nitride present natural directions for future research.

\begin{acknowledgments}
The authors wish to thank I. L. Aleiner, M. O. Goerbig, and A. R. Wieteska for useful conversations.  Work is supported by a Research Corporation Cottrell College Science Award and the Bard Summer Research Institute.  A portion of this work was performed at the National High Magnetic Field Laboratory, which is supported by National Science Foundation Cooperative Agreement No. DMR-1157490 and the State of Florida.

\end{acknowledgments}

\end{document}